\documentclass[conference]{IEEEtran}
\IEEEoverridecommandlockouts
\usepackage{cite}
\usepackage{amsmath,amssymb,amsfonts}
\usepackage{algorithm}
\usepackage{graphicx}
\usepackage{textcomp}
\usepackage{xcolor}
\usepackage{algpseudocode}
\usepackage{amsfonts}
\usepackage{booktabs}
\usepackage{multirow}
\usepackage{booktabs}
\usepackage{subcaption} 
\usepackage{tabularx}
\usepackage{placeins}

\newcommand{\blue}[1]{\textcolor{blue}{#1}}

\def\BibTeX{{\rm B\kern-.05em{\sc i\kern-.025em b}\kern-.08em
    T\kern-.1667em\lower.7ex\hbox{E}\kern-.125emX}}

\begin{document}

\title{Selection of Optimal Number and Location of PMUs for CNN Based Fault Location and Identification}

\author{\IEEEauthorblockN{1\textsuperscript{st} Khalid Daud Khattak}
\IEEEauthorblockA{\textit{Lane Department of Computer Science}\\ 
\textit{and Electrical Engineering,}\\
\textit{West Virginia University,}\\
Morgantown, WV USA \\
kk0041@mix.wvu.edu}

\and
\IEEEauthorblockN{2\textsuperscript{nd} Muhammad A. Choudhry}
\IEEEauthorblockA{\textit{Lane Department of Computer Science}\\ 
\textit{and Electrical Engineering}\\ 
\textit{West Virginia University,}\\
Morgantown, WV USA}}

\maketitle

\begin{abstract}
In this paper a data‐driven Forward Selection with Neighborhood Refinement (FSNR) algorithm is presented to determine the number and placement of Phasor Measurement Units (PMUs) for maximizing deep‐learning–based fault diagnosis performance. Candidate PMU locations are ranked via a cross‐validated Support Vector Machine (SVM) classifier, and each selection is refined through local neighborhood exploration to produce a near‐optimal sensor set. The resulting PMU subset is then supplied to a 1D Convolutional Neural Network (CNN) for faulted‐line localization and fault‐type classification from time‐series measurements. Evaluation on modified IEEE-34 and IEEE-123 bus systems demonstrates that the proposed FSNR-SVM method identifies a minimal PMU configuration that achieves the best overall CNN performance, attaining over 
96\% accuracy in fault location and over 99\% accuracy in fault type classification on the IEEE-34 system, and approximately 94\% accuracy in fault location and around 99.8\% accuracy in fault type classification on the IEEE-123 system. 
\end{abstract}

\begin{IEEEkeywords}
Convolutional neural networks, Fault location, Phasor measurement units, Random forests, Support vector machines. 
\end{IEEEkeywords}  

\section{Introduction}
The deployment of Phasor Measurement Units (PMUs) has become increasingly practical in power distribution networks (PDNs) since the development of cost-effective micro-PMUs ($\mu$PMUs) and distribution level PMUs (D-PMUs). This is particularly significant for modern smart grid applications, which rely on real-time measurements to facilitate faster, more reliable, and autonomous decision-making. The time-synchronized voltage and current measurements provided by such devices can be useful for fault location and fault type identification (FLI) \cite{10106028}. Despite these advances, PDNs often comprise a large number of nodes, making it infeasible to equip every node with a PMU. This results in partial system observability, which poses a critical challenge for FLI. The integration of distributed energy resources (DERs), such as photovoltaics (PV) and wind turbine generators (WTGs), further complicate FLI due to the introduction of bidirectional power flow and dynamic system behavior.
To address these challenges, various methods have been proposed in the literature, with a growing emphasis on data-driven approaches that use artificial intelligence (AI). In particular, deep learning models such as convolutional neural networks (CNNs) and Residual Networks (ResNets) have shown significant promise \cite{9650703, 8718345, 10454243}. These models are capable of learning complex, hidden patterns in high-dimensional data that traditional machine learning methods may fail to detect. To accommodate PMU data, 1D versions of deep neural network models can be used. This not only eliminates the need for 2D data transformations but also offers computational advantages as 1D models are more efficient to train and often require fewer resources \cite{KIRANYAZ2021107398}. Deep Neural Network (DNN) performance depends on effective data preprocessing, appropriate choice of network depth, convolutional filter sizes, learning rate tuning, and nonlinear activation functions. Equally important is feature selection—in fault location and identification tasks, this entails determining the minimal number and best placement of PMUs to provide the most informative data for training deep learning models.    

\subsection {Related Work}
 In~\cite{8892526}, the authors have proposed a PMU placement algorithm for power distribution networks (PDNs) that minimizes \textit{observability loss} by analyzing the structure of the system admittance matrix. Specifically, the objective is to minimize the infinity norm of the singular values of the matrix \( Y_{au} Y_{uu}^{-1} \), expressed as \( \left\| \sigma\left(Y_{au} Y_{uu}^{-1} \right) \right\|_{\infty} \), where \( \sigma \) denotes the vector of singular values, \( Y_{uu} \) is the admittance submatrix corresponding to unmonitored buses, and \( Y_{au} \) captures the coupling between monitored and unmonitored buses. The algorithm adopts a greedy selection strategy, iteratively placing PMUs at candidate locations that yield the greatest reduction in the objective function. The method, however, does not provide the optimal number of PMUs to use but only gives the best locations to place a certain number of PMUs in the PDN. 
 A sensitivity-based PMU placement strategy is described in \cite{https://doi.org/10.1049/gtd2.12055}. Although the method takes into account the variations in fault resistance and fault distance, it does not consider the impact of DERs in the PDN. 
 In~\cite{9774839}, the concept of Usable Zero-Injection Phases (UZIPs) is introduced to enable indirect observability of unmeasured nodes through a single $\mu$PMU. The proposed method incorporates key distribution system features such as unbalanced phase configurations, DERs, and network reconfiguration into the PMU placement strategy. Simulation results are presented for the IEEE-13, 34, and 123-bus test systems. However, a key limitation of this approach is its reliance on the presence of zero-injection buses (ZIPs), which may not be consistently available in practical systems due to load variability and incomplete system modeling. Furthermore, the reported placement of 21 $\mu$PMUs in the IEEE-34 bus system is relatively high, raising concerns about scalability and cost-effectiveness for larger networks.
 In~\cite{8718345}, a greedy search algorithm has been integrated with a CNN to determine optimal PMU placement in PDNs by minimizing the cross-entropy loss of faulted line predictions. This approach is notable for directly leveraging deep learning to guide placement decisions, potentially capturing complex spatiotemporal patterns in PMU data. However, the method can be computationally intensive, particularly due to the repeated training of the CNN during the greedy search process. As a result, a full exploration of all PMUs may not be feasible. Furthermore, the greedy nature of the algorithm may lead to suboptimal placements, as it does not guarantee a globally optimal solution.

 \subsection{Motivation and Contributions}
 From the literature, various methods have been proposed for optimal PMU placement in PDNs. However, most are not tailored for integration with DNN architectures. Since input data quality and configuration strongly influence DNN performance, it is essential to determine both the number and placement of PMUs for accurate fault location and classification. Practical deployment also requires balancing accuracy with the cost–benefit of PMU installation.

To address this, we propose a data-driven method that leverages a conventional machine learning model—specifically a Support Vector Machine (SVM)—to enable rapid training and evaluation across multiple sensor placement configurations. Our approach employs a forward greedy search algorithm, which incrementally selects PMU locations based on classification accuracy. SVMs are particularly suitable for this task due to their effectiveness in high-dimensional feature spaces and robustness when trained on relatively small datasets. SVM based feature selection methods have been explored in \cite{sanz_svm-rfe_2018, MALDONADO20092208}. 

To reduce the impact of globally suboptimal selection, possible in greedy algorithms, we introduce a neighborhood refinement mechanism. This step involves revisiting the second most recently added PMU location and evaluating alternative placements based on their contribution to classification performance.  

\section{Proposed Search Algorithm}
 Since the type and location of faults are highly sensitive to voltage and current variations observed both pre-fault and during-fault at different buses, we incorporate these dynamics into our feature set. Specifically, we consider:

\begin{itemize}
    \item The difference of pre-fault and during fault three symmetrical components (zero, positive, and negative sequences) of both voltages and currents measured at the substation,

\begin{equation}
\label{eq1}
\Delta \mathbf{V}_{0,1,2} = \left[ \mathbf{V}_{0,1,2}^{\text{pre-fault}} - \mathbf{V}_{0,1,2}^{\text{during fault}} \right]_{\text{substation}},
\end{equation}

\begin{equation}
\label{eq2}
\Delta \mathbf{I}_{0,1,2} = \left[ \mathbf{I}_{0,1,2}^{\text{pre-fault}} - \mathbf{I}_{0,1,2}^{\text{during fault}} \right]_{\text{substation}}.
\end{equation}

\item Only the difference of pre-fault and during fault three symmetrical components of voltages measured at other buses in the distribution network,

\begin{equation}
\label{eq3}
\Delta \mathbf{V}_{0,1,2} = \left[ \mathbf{V}_{0,1,2}^{\text{pre-fault}} - \mathbf{V}_{0,1,2}^{\text{during fault}} \right]_{\text{node }i}.
\end{equation}

\end{itemize}

The training data is generated by simulating all types of faults at the midpoint of each line in the PDN, with a fixed fault resistance. Fault scenarios are created for different hours over a 24-hour period, capturing variations in voltage due to changes in DERs outputs across time. The output labels consist of the faulted line and fault type, both of which are encoded as numerical categories for machine learning.

Given a training set \( \{(\mathbf{x}_i, y_i)\}_{i=1}^n \), where \( y_i \in \{-1, +1\} \), the soft-margin SVM seeks a hyperplane in a feature space (via mapping \( \phi(\cdot) \)) by solving:

\begin{align}
\min_{\mathbf{w}, b, \xi} \quad & \frac{1}{2} \|\mathbf{w}\|^2 + C \sum_{i=1}^{n} \xi_i, \nonumber \\
\text{subject to } \quad & y_i (\mathbf{w}^\top \phi(\mathbf{x}_i) + b) \geq 1 - \xi_i,\quad \xi_i \geq 0.
\end{align}

where $\mathbf{w}$ is the weight vector, $\xi$ represents the slack variable, and $C$ represents the regularization parameter. SVMs leverage kernel functions to handle non-linear separability. The widely used radial basis function (RBF) kernel is:

\[
K(\mathbf{x}_i, \mathbf{x}_j) = \exp(-\gamma \|\mathbf{x}_i - \mathbf{x}_j\|^2).
\]

where $\gamma $ is RBF kernel width parameter. The decision function for a new data point \( \mathbf{x} \) is:

\[
f(\mathbf{x}) = \sum_{i=1}^{n} \alpha_i y_i K(\mathbf{x}_i, \mathbf{x}) + b,
\]
where \( \alpha_i \) are the learned Lagrange multipliers.

For multi-class classification, the \textit{one-vs-rest} strategy trains \( K \) binary classifiers. For class \( k \), the labels are:

\[
y_i^{(k)} =
\begin{cases}
+1 & \text{if } y_i = k, \\
-1 & \text{otherwise}.
\end{cases}
\]
The final prediction is made as:

\[
\hat{y} = \arg\max_{k \in \{1, \dots, K\}} f_k(\mathbf{x}).
\]

 Let \(X\) include the $d$ concatenated three-phase measurements of Eqs. \ref{eq1}, \ref{eq2}, and \ref{eq3}, of the $(d-12)/6  +1 $ PMUs. The proposed Forward Selection with Neighborhood Refinement (FSNR) with SVM classifier is given in Algorithm \ref{alg:forward-selection}.

\begin{algorithm}[htbp]
\caption{Greedy Forward Selection with Neighborhood Refinement (FSNR) with SVM classfier}
\label{alg:forward-selection}
\begin{algorithmic}[1]
\Require Candidate PMUs $N$, budget $m$, SVM hyperparameters $(\gamma, C)$, CV folds $cv$
\Ensure Selected set $Q$
\State $Q\gets\{\text{substation volts},\,\text{substation currents}\}$ \Comment{always--keep measurements}
\State $\textit{Available}\gets N\setminus Q$
\State \textbf{Preprocess:} normalize feature vectors of every $p\in\textit{Available}$
\State $\textit{BaseScore}\gets\text{score}(Q)$
\While{$|Q|<m$}
    \State $p^*\gets\displaystyle\arg\max\,\text{score}(Q\cup\{p\})$ \Comment{$p\in\textit{Available}$}
    \State $Q\gets Q\cup\{p^*\}$;\; $\textit{Available}\gets\textit{Available}\setminus\{p^*\}$
    \If{$|Q|>3$ \textbf{and} $\text{score}(Q)>\textit{BaseScore}$}
        \State $p_{\text{prev}}\gets$ node added in the previous iteration
        \State $v^*\gets\displaystyle\arg\max\,\text{score}\!\bigl((Q\setminus\{p_{\text{prev}}\})\cup\{v\}\bigr)$ \Comment{$v\in\mathcal N(p_{\text{prev}})\cap\textit{Available}$}
        \If{$\text{score}((Q\setminus\{p_{\text{prev}}\})\cup\{v^*\})>\text{score}(Q)$}
            \State $Q\gets(Q\setminus\{p_{\text{prev}}\})\cup\{v^*\}$
            \State adjust $\textit{Available}$ accordingly
        \EndIf
    \EndIf
    \State $\textit{BaseScore}\gets\text{score}(Q)$
\EndWhile
\State \Return $Q$
\end{algorithmic}
\end{algorithm}

Initially, the baseline accuracy of the SVM classifier is computed using measurements from the substation alone. PMUs are then iteratively added, one at a time, on the basis of the improvement they contribute to classification accuracy. At each iteration, the PMU whose inclusion yields the greatest increase in accuracy is added to the candidate set.

To mitigate the risk of making locally optimal (but globally suboptimal) choices during this greedy selection process, we introduce a neighborhood refinement step. Whenever a new sensor increases the classification accuracy, we also consider replacing the previously added PMU with each of its four immediate upstream and downstream neighbors. If any replacement yields a further accuracy improvement, we adopt it. In this way, the order in which PMUs enter the set directly reflects their incremental value for fault location and type classification. Finally, we consider the best PMU count to use as the point where adding more sensors no longer leads to significant accuracy gains.

\section{Data Preparation and Implementation Details}
\subsection{Test Systems}
We use the IEEE-34 and IEEE-123 bus systems \cite{IEEEPES_TestFeeders} to evaluate the performance of the proposed PMUs placement method. The base cases have been modified to include the impacts of DERs. These models have been implemented using OpenDSS software in conjunction with Py-Dss Interface library\cite{radatz2024bdgd2opendss}. Renewable DERs are operated along different operational curves to reflect real-time variation in generation. Table \ref{tab:DER_rows} gives the details of the DER placement in the test systems.

\begin{table}[htbp]
\caption{DER Placements and Ratings for the IEEE 34-Bus and IEEE 123-Bus Test Feeders}
\centering
\scriptsize           % shrink font to fit one column
\renewcommand{\arraystretch}{1.1}
\begin{tabular}{@{}|c|c|c|c|c|@{}}
\hline
\textbf{System} & \textbf{Type} & \textbf{ID} & \textbf{Location} & \textbf{Rating} \\ \hline
\multirow{3}{*}{IEEE-34} 
  & PV  & PV1  & 840 via busPV1 & 240 kW \\ \cline{2-5}
  & WTG & WTG1 & 844 via buswind1 & 250 kW @ 0.96 PF \\ \cline{2-5}
  & DG  & DG1  & 890 & 180 kW @ 0.98 PF \\ \hline
\multirow{6}{*}{IEEE-123} 
  & PV  & PV1  & 35 via PVgen1  & 150 kW \\ \cline{2-5}
  & PV  & PV2  & 78 via PVgen2  & 100 kW \\ \cline{2-5}
  & PV  & PV3  & 64 via PVgen3  &  80 kW \\ \cline{2-5}
  & WTG & WTG1 & 48 via windgen1  & 200 kW @ 0.98 PF \\ \cline{2-5}
  & WTG & WTG2 & 95 via windgen2  & 100 kW @ 0.98 PF \\ \cline{2-5}
  & WTG & WTG3 & 108 via windgen3 & 100 kW @ 0.98 PF \\ \hline
  
\end{tabular}
\label{tab:DER_rows}
\end{table}

\subsection{PMU Placement Data and Strategies}
The datasets used with the proposed algorithm for the placement of PMUs are prepared according to \eqref{eq1}, \eqref{eq2} and \eqref{eq3}. Data is recorded every two hours for the IEEE-34 bus system and every four hours for the IEEE-123 bus system with a fault resistance of $0.001 \Omega$ and a ground resistance of $1\Omega$. Let $Q\subseteq N$ denote the current set of candidate PMU locations and
$\mathbf X^{(Q)}\!\in\!\mathbb{R}^{n\times 6|Q|}$ the corresponding
fault–sample matrix (first twelve columns for real and imaginary components of voltage and current measurements at the substation and six columns per bus for the remaining buses). To show the effectiveness of our proposed FSNR with SVM classifier, we compare it with the following PMU selection approaches.
\begin{itemize}
\item \textbf{SVM based Forward Search (FS) without Neighborhood Refinement}
    \item \textbf{Random Forest Classifier based FSNR}
    \item \textbf{Correlation distance based FSNR:}
The correlation distance scalar score $D(Q)$ is evaluated in four
steps:

\paragraph*{1) Column‐wise $\ell_{2}$ normalization}
\begin{equation}
\widetilde{\mathbf X}^{(Q)}
  \;=\;\text{norm}_{2}\!\bigl(\mathbf X^{(Q)}\bigr),
\qquad
\bigl\|\widetilde x_{\bullet r}\bigr\|_{2}=1
\;\;\forall\,r .
\label{eq:norm2}
\end{equation}

\paragraph*{2) Sequence-magnitude extraction}
From every contiguous block of six columns we form the
zero-, positive- and negative-sequence magnitudes  
$\mathbf X_{0}^{(Q)},\mathbf X_{1}^{(Q)},\mathbf X_{2}^{(Q)}
\in\mathbb{R}^{n\times |Q|}$ by
\begin{align}
X_{k}^{(Q)}(s,i) &=
\sqrt{
  \bigl(\widetilde x_{s,\,2k+6i}\bigr)^{2}
  + \bigl(\widetilde x_{s,\,2k+6i+1}\bigr)^{2}
},
\notag \\
&\quad k \in \{0,1,2\}.
\label{eq:magnitudes}
\end{align}
\paragraph*{3) Vertical concatenation}
The three blocks are stacked to obtain
\begin{align}
\mathbf Z^{(Q)} 
&:= 
\left(
\begin{bmatrix}
(\mathbf X_{0}^{(Q)})^{\top} \quad
(\mathbf X_{1}^{(Q)})^{\top} \quad
(\mathbf X_{2}^{(Q)})^{\top}
\end{bmatrix}
\right)^{\top} \notag \\
&\in \mathbb{R}^{3n \times |Q|},
\quad
\mathbf u_i := \mathbf Z^{(Q)}_{\bullet i}.
\label{eq:stack}
\end{align}

\paragraph*{4) Sum of absolute correlation distances}
With $\bar{\mathbf u}_{i}$ the sample mean of $\mathbf u_{i}$, define  
the Pearson correlation of every bus pair,
\begin{equation}
\rho_{ij} \;=\;
\frac{(\mathbf u_{i}-\bar{\mathbf u}_{i})^{\!\top}
      (\mathbf u_{j}-\bar{\mathbf u}_{j})}
     {\bigl\|\,\mathbf u_{i}-\bar{\mathbf u}_{i}\bigr\|_{2}\;
      \bigl\|\,\mathbf u_{j}-\bar{\mathbf u}_{j}\bigr\|_{2}},
\label{eq:rho_ij}
\end{equation}
and the overall diversity score
\begin{equation}
D(Q) \;=\;
\sum_{i<j}
\Bigl(1-\lvert\rho_{ij}\rvert\Bigr)
\;\in[0, |Q|(|Q|-1)/2].
\label{eq:DQ}
\end{equation}

\noindent
Because $1-\lvert\rho_{ij}\rvert$ equals~$0$ for perfectly (positively or negatively) correlated buses and~$1$ for uncorrelated buses, maximizing $D(Q)$ with FSNR algorithm helps select PMU locations whose phasor magnitudes
are as linearly independent as possible across \emph{all} three symmetrical components.
    \item \textbf{Admittance matrix based:} PMU placement algorithm suggested in \cite{8892526}.
\end{itemize}

\subsection{1D CNN for Fault Location and Fault Type Identification}
We use a 1D CNN model inspired from the VGG-11 architecture in \cite{Simonyan2015VGG}. The details of the 1D CNN model are provided in Table \ref{tab:cnn-summary}.

% \begin{table}[htbp]
% \caption{Model Summary 1D CNN for IEEE-34 and IEEE-123 Bus Systems}
% \label{tab:cnn-summary}
% \centering
% \setlength{\tabcolsep}{3pt} % tighten column spacing
% \renewcommand{\arraystretch}{1.15} % compact row height
% \begin{tabular}{|p{3.85cm}|c|r|c|r|}
% \hline
% \multirow{2}{*}{\textbf{Module/Layer}} & \multicolumn{2}{c|}{\textbf{IEEE-34}} & \multicolumn{2}{c|}{\textbf{IEEE-123}} \\ \cline{2-5}
%  & S-Len & \#Params & S-Len & \#Params \\ \hline
% \multicolumn{5}{|l|}{\textit{Feature Extraction}} \\ \hline
% InputConv: Conv1d 30$\rightarrow$128, k=3, BN, ReLU & 36 & 12,032 & 60 & 12,032 \\ \hline
% 2$\times$Conv1d(128), BN, GELU+MP & 18 & 99,072 & 30 & 99,072 \\ \hline
% 3$\times$Conv1d(256), BN, GELU+MP & 9 & 493,824 & 15 & 493,824 \\ \hline
% 3$\times$Conv1d(512), BN, GELU+MP & 4 & 1,970,688 & 7 & 1,970,688 \\ \hline
% Flatten & 2048 & -- & 3584 & -- \\ \hline
% \multicolumn{5}{|l|}{\textbf{Classifier Head}} \\ \hline
% FC 2048$\rightarrow$512, GELU, Dropout & 512 & 1,049,088 & -- & 1,835,520 \\ \hline
% FC 512$\rightarrow$256, GELU, Dropout & 256 & 131,328 & -- & 131,328 \\ \hline
% Output 1 (fault type): 256$\rightarrow$11 & 11 & 2,827 & 11 & 2,827 \\ \hline
% Output 2 (fault loc): 256$\rightarrow n_{\text{loc}}$ & 34 & 8,738 & 118 & 30,326 \\ \hline
% \textbf{Total} & -- & \textbf{3,767,597} & -- & \textbf{4,575,617} \\ \hline
% \end{tabular}
% \label{tab:model_summary}
% \vspace{2pt}
% \footnotesize{$n_{\text{loc}}$: number of fault--location classes; BN: Batch Normalization; MP: MaxPool.}
% \end{table}

\begin{table}[htbp]
\caption{1D CNN Summary: Shared Convolutional Parameters and System-Specific Classifier Parameters}
\label{tab:cnn-summary}
\centering
\small
\setlength{\tabcolsep}{3pt}
\renewcommand{\arraystretch}{1.12}

% ---------- Shared Feature Extraction ----------
\begin{tabular}{|p{4.5cm}|c|c|r|}
\hline
\multicolumn{4}{|l|}{\textbf{Shared Feature Extraction}} \\ \hline
\textbf{Module/Layer} & \textbf{Len-34} & \textbf{Len-123} & \textbf{Params} \\ \hline
InputConv: Conv1d 30$\rightarrow$128, k=3, BN, ReLU & 36 & 60 & 12{,}032 \\ \hline
2$\times$Conv1d(128), BN, GELU + MP & 18 & 30 & 99{,}072 \\ \hline
3$\times$Conv1d(256), BN, GELU + MP & 9 & 15 & 493{,}824 \\ \hline
3$\times$Conv1d(512), BN, GELU + MP & 4 & 7 & 1{,}970{,}688 \\ \hline
Flatten & 2048 & 3584 & -- \\ \hline
\end{tabular}

\vspace{1pt}

% ---------- Classifier Head (System-Specific) ----------
\begin{tabular}{|p{4.5cm}|p{1.8cm}|p{1.8cm}|}
\hline
\multicolumn{3}{|l|}{\textbf{Classifier Head (system-specific parameters)}} \\ \hline
\textbf{Module/Layer} &
\textbf{Params-34} & \textbf{Params-123} 
 \\ \hline
FC 2048$\rightarrow$512, GELU, Dropout & 1{,}049{,}088 & 1{,}835{,}520 \\ \hline
FC 512$\rightarrow$256, GELU, Dropout & 131{,}328 & 131{,}328 \\ \hline
Output 1 (fault type): 256$\rightarrow$11 & 2{,}827 & 2{,}827 \\ \hline
Output 2 (fault loc): 256$\rightarrow n_{\text{loc}}$ & 8{,}738 & 30{,}326 \\ \hline
\textbf{Total} & \textbf{3{,}767{,}597} & \textbf{4{,}575{,}617} \\ \hline
\end{tabular}

\vspace{2pt}
\footnotesize{Len: sequence length at that stage;\; $n_{\text{loc}}$: number of fault--location classes;\; BN: Batch Normalization;\; MP: MaxPool.}
\end{table}

In Table~\ref{tab:cnn-summary}, $n_{\rm loc}$ is 34 for the IEEE-34 bus system and 118 for IEEE-123.  We train our CNN using the standard multiclass cross-entropy loss,
\[
\mathcal{L}_{\rm CE}
= -\frac{1}{N}\sum_{i=1}^N
\log\!\Biggl(\frac{\exp\bigl(z_{\,t^{(i)}}^{(i)}\bigr)}
{\sum_{k=1}^K\exp\bigl(z_{\,k}^{(i)}\bigr)}\Biggr)\,,
\]
where $N$ is the batch size and $K$ is the number of output classes. For each example $i$, the network produces logits $z^{(i)}_1,\dots,z^{(i)}_K$ and $t^{(i)}\in\{1,\dots,K\}$  denotes its true class index. The softmax function in the logarithm converts these logits into a probability distribution over the $K$ classes. The CNN model was run for 400 epochs for each test case with the Adam optimizer, using a weight decay of $10^{-6}$ and an initial learning rate of $5 \times 10^{-4}$. After each epoch, the learning rate is exponentially decayed by a factor of $\gamma=0.987$ via PyTorch’s \texttt{ExponentialLR} scheduler:
\[
  \eta_{t+1} = \gamma\,\eta_t
  \quad\Longrightarrow\quad
  \eta_t = \eta_0\,\gamma^t,
\]
where $t$ is the epoch index. This gradual decay reduces the step size over time, helping the model converge more smoothly.  

\subsection{Datasets for 1D CNN Model}
The datasets consist of 60,000 samples for the IEEE-34 bus system and 80,000 samples for the IEEE-123 bus system. The datasets are split in the ratio 70-15-15 for training, validation, and testing respectively. Faults have been simulated by moving the fault point randomly along a selected line. Fault resistance has been randomly varied in the range $0.01\Omega - 10 \Omega$, while the ground resistance has been randomly selected from the set $\left\{ 1\,\Omega,\, 5\,\Omega,\, 10\,\Omega, 20\,\Omega\right\}$. Fault location is done by identifying the faulted line, and fault‐type classification is achieved by identifying one of eleven classes—three LG (single line‐to‐ground), three LL (line‐to‐line), three LLG (double line‐to‐ground), one LLL (three‐phase), or one LLLG (three‐phase‐to‐ground) fault.
 The output labels for fault location and fault type classification are one-hot encoded. Table \ref{tab:fault_counts} gives the distribution of the datasets for IEEE-34 and IEEE-123 bus systems according to the different fault types.

\begin{table}[!ht]
\centering
\caption{Fault sample counts by type and split}
\label{tab:fault_counts}
\small
\setlength\tabcolsep{2pt}  % tighten up inter‐column spacing
\begin{tabular}{@{}l rrr|rrr@{}}
\toprule
\textbf{Fault} 
  & \multicolumn{3}{c|}{\textbf{34 Bus}} 
  & \multicolumn{3}{c}{\textbf{123 Bus}} \\
\cmidrule(lr){2-4} \cmidrule(lr){5-7}
\textbf{Type} 
  & \textbf{Train.} & \textbf{Val.} & \textbf{Test.} 
  & \textbf{Train.} & \textbf{Val.} & \textbf{Test.} \\
\midrule
AG    & 4349   & 941   & 926   & 6528   & 1418   & 1407   \\
BG    & 4304   & 918   & 882   & 5669   & 1186   & 1225   \\
CG    & 3742   & 754   & 790   & 6348   & 1342   & 1358   \\
AB    & 3611   & 803   & 740   & 4710   & 1016   &  920   \\
BC    & 3754   & 796   & 814   & 4680   &  982   & 1015   \\
AC    & 3695   & 774   & 832   & 4817   &  998   & 1077   \\
ABG   & 3759   & 785   & 848   & 4773   & 1032   &  990   \\
BCG   & 3669   & 798   & 795   & 4625   & 1008   &  985   \\
ACG   & 3729   & 790   & 778   & 4698   & 1043   &  991   \\
ABC   & 3650   & 809   & 745   & 4531   & 1002   & 1028   \\
ABCG  & 3738   & 832   & 850   & 4621   &  973   & 1004   \\
\midrule
\textbf{Total}
      & 42000  & 9000  & 9000  & 56000  & 12000  & 12000  \\
\bottomrule
\end{tabular}
\end{table}

The data is sampled at 60 Hz as per IEEE standard for PMUs \cite{IEEE_C37_118_1_2011}. We collect 15 pre-fault and 15 during fault samples.

\section{Results}
\begin{figure*}[!ht]
    \centering
    % First subplot
    \begin{subfigure}[t]{0.329\textwidth}
        \centering
        \includegraphics[width=\textwidth]{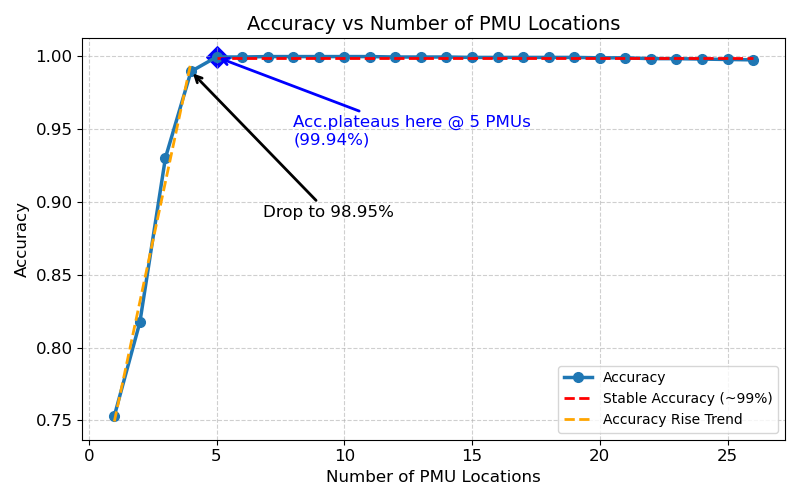}
        \caption{}
    \end{subfigure}
    \hfill
    % Second subplot
    \begin{subfigure}[t]{0.329\textwidth}
        \centering
        \includegraphics[width=\textwidth]{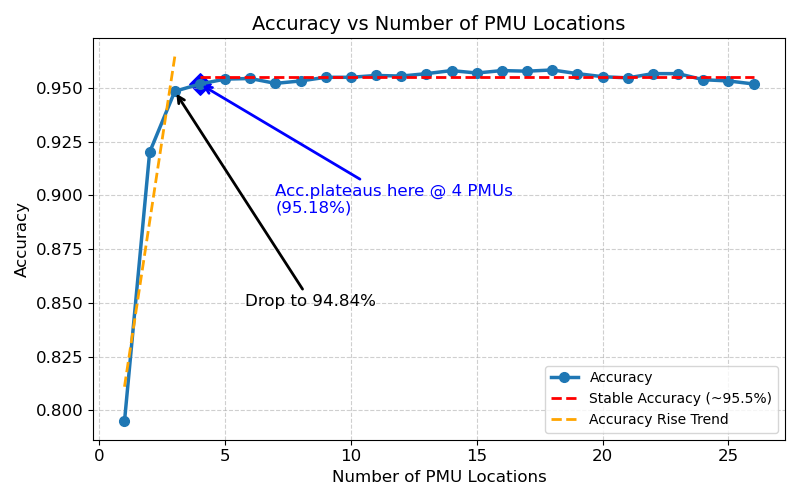}
        \caption{}
    \end{subfigure}
    \hfill
    %Third subplot
    \begin{subfigure}[t]{0.329\textwidth}
        \centering
        \includegraphics[width=\textwidth]{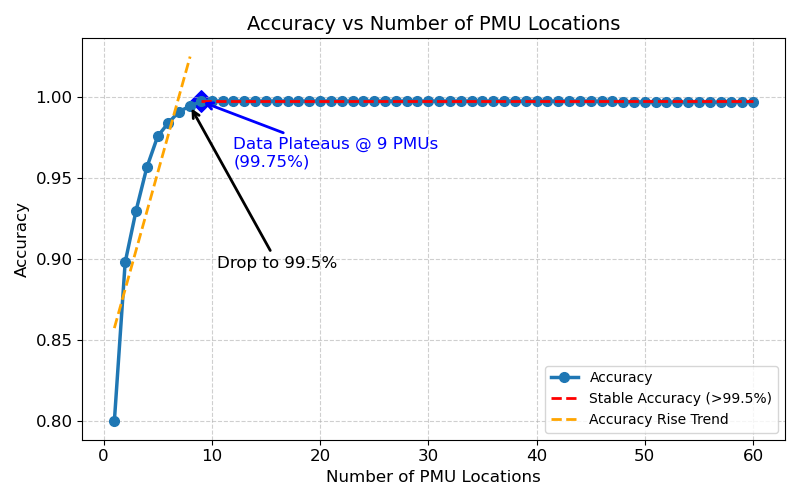}
        \caption{}
    \end{subfigure}
    \hfill

    \caption{Accuracy vs. Number of PMUs: (a) IEEE-34 Bus System -- FSNR with SVM, (b) IEEE-34 Bus System -- FSNR with Random Forest, and (c) IEEE-123 Bus System -- FSNR with SVM.}
    \label{fig:ACCvsNoPMUplot}
\end{figure*}

Hyperparameters for the SVM classifier were tuned by a full‐grid search over $C$ and $\gamma$ using all available PMU locations. In case of the IEEE-34 system, we found the best performance at $\gamma$=500, $C$=1500, whereas on IEEE-123 both parameters settled at 500. Fig. \ref{fig:ACCvsNoPMUplot} plots classification accuracy as a function of the number of PMUs: for IEEE-34 the curve flattens with five PMUs, and for IEEE-123 it plateaus at nine.

Fig.~\ref{fig:ACCvsNoPMUplot}b repeats the experiment using a Random Forest classifier for the IEEE-34 bus system. Its accuracy plateaus after about four PMUs but with higher variance and a lower ceiling (\(\approx95\%\)) than the SVM (\(\approx99.9\%\)). In case of the IEEE-123 bus case, the Random Forest becomes prohibitively slow as more PMUs are added, so we only evaluated up to twenty PMU locations. However, beyond nine PMUs, adding more did not yield any significant improvement.

Accordingly, in all subsequent comparisons we fix the PMU counts to those determined by FSNR-SVM (five on IEEE-34, nine on IEEE-123). Table \ref{tab:bus_locations} lists the final PMU sets for each placement algorithm. Notably, some buses recur across different selection methods — for instance, buses \lq848\rq\  and \lq{busPV1}\rq\ are chosen by every algorithm in the IEEE-34 system, while buses \lq83\rq\, \lq95\rq\, and \lq151\rq\ appear in three out of four methods for IEEE-123—highlighting their fault-discriminative importance. 

\begin{table}[ht]
\centering
\caption{PMU Placement Bus Locations for IEEE‐34 and IEEE‐123}
\label{tab:bus_locations}
% \begin{tabularx}{\linewidth}{@{}l >{\hsize=0.6\hsize}X >{\hsize=1.4\hsize}X@{}}
\begin{tabularx}{\linewidth}{@{}p{0.11\linewidth} p{0.29\linewidth} p{0.5\linewidth}@{}}
\toprule
\textbf{Test System} & \textbf{Method}                 & \textbf{Bus Locations}                                   \\
\midrule
\multirow{4}{*}{IEEE‐34}
 & FSNR-SVM            & 800, 848, busPV1, 850, 854                               \\
& FS-SVM & 800, 848, busPV1, 850, 828
\\

 & FSNR-Random Forest  & 800, 848, busPV1, buswind1, 854                          \\
 & Correlation Distance     & 800, 848, 890, 850, busPV1, 862                          \\
 & Admittance Matrix\cite{8892526}        & 800, 848, busPV1, 830, 812                               \\
\midrule
\multirow{4}{*}{IEEE‐123}
 & FSNR-SVM & 149, 83, 151, 300, 95, 250, 66, 56, 450                  \\
&FS-SVM & Same as FSNR-SVM
                \\
 & FSNR-Random Forest  & 149, 83, 51, 450, 95, 65, 29, 300, 151                   \\
 & Correlation Distance     & 149, 95, 151, 93, 51, 83, 250, 91, 50                    \\
 & Admittance Matrix\cite{8892526}  & 149, 81, 99, 30, 61, 56, 50, 89, 48           \\
\bottomrule
\end{tabularx}
\end{table}

\begin{figure}[ht!]
  \centering
  % adjust width as needed: \linewidth, 0.8\linewidth, etc.
  \includegraphics[width=0.9\linewidth]{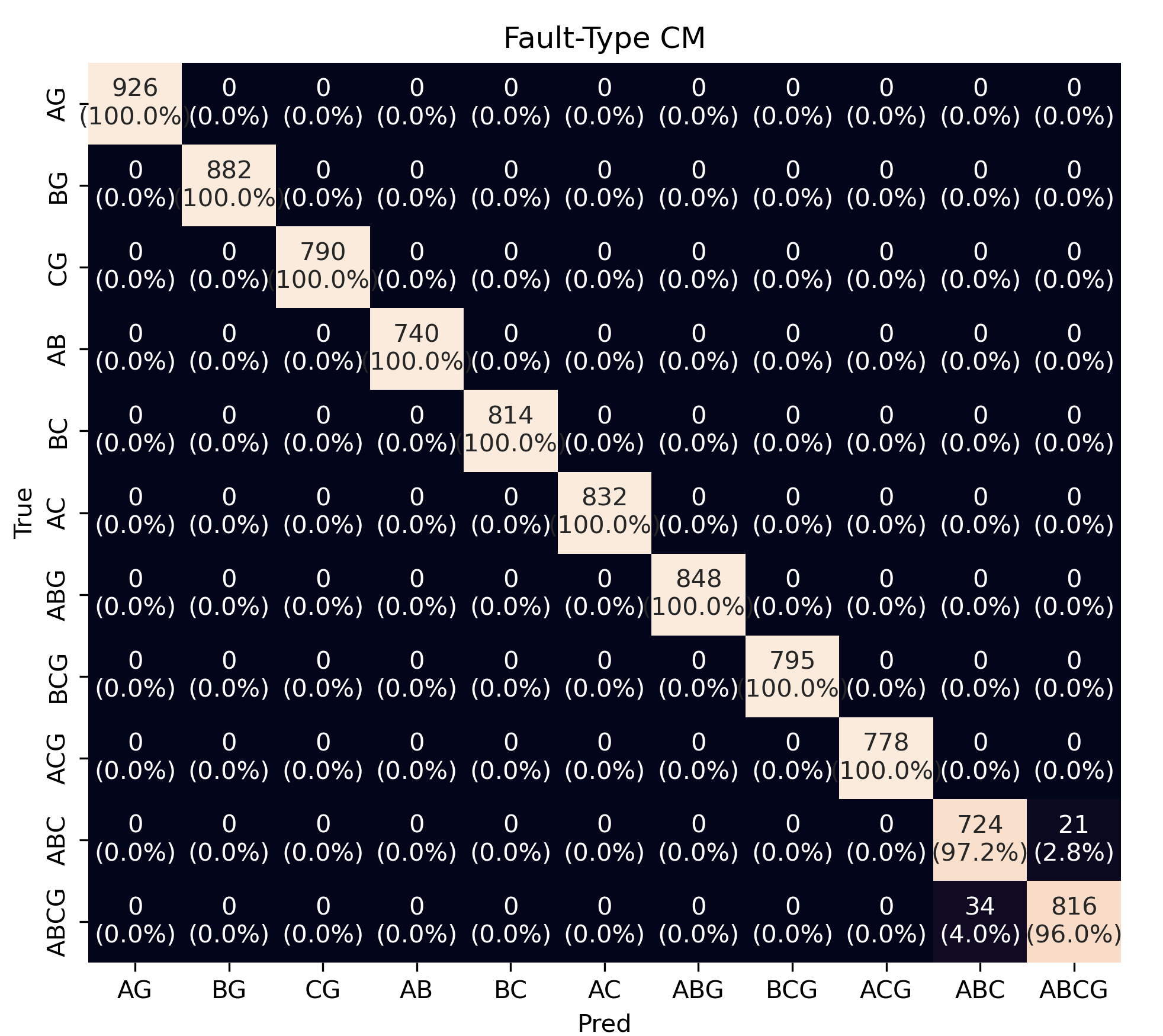}
  \caption{Confusion matrix of fault classification for IEEE-34 Bus System with PMUs selected using FSNR-SVM.}
  \label{fig:IEEE-34-confusion-mat}
\end{figure}

The results of the 1D CNN model with different PMU placement strategies are presented in Table \ref{tab:perf}, evaluated using standard metrics such as accuracy, precision, recall, F1 score, and specificity. For the IEEE-34 bus system, FSNR-SVM based placement achieves the best fault location performance, yielding nearly a $1\%$ accuracy gain over non-SVM methods. The admittance matrix-based algorithm proposed in [6] produces the best fault type classification, though its margin over FSNR-SVM is minimal. Misclassifications are primarily observed between the ABC and ABCG fault types. All other fault categories were successfully identified with 100\% accuracy. This can be seen in the confusion matrix for fault type classification, shown in Fig. \ref{fig:IEEE-34-confusion-mat}. In this case of IEEE-34 bus system, using the SVM classifier with forward search (without neighborhood refinement) leads to a different placement of just one PMU, yet this small change still affects CNN performance.  

\begin{figure}[!htb]
  \centering
  % adjust width as needed: \linewidth, 0.8\linewidth, etc.
  \includegraphics[width=0.9\linewidth]{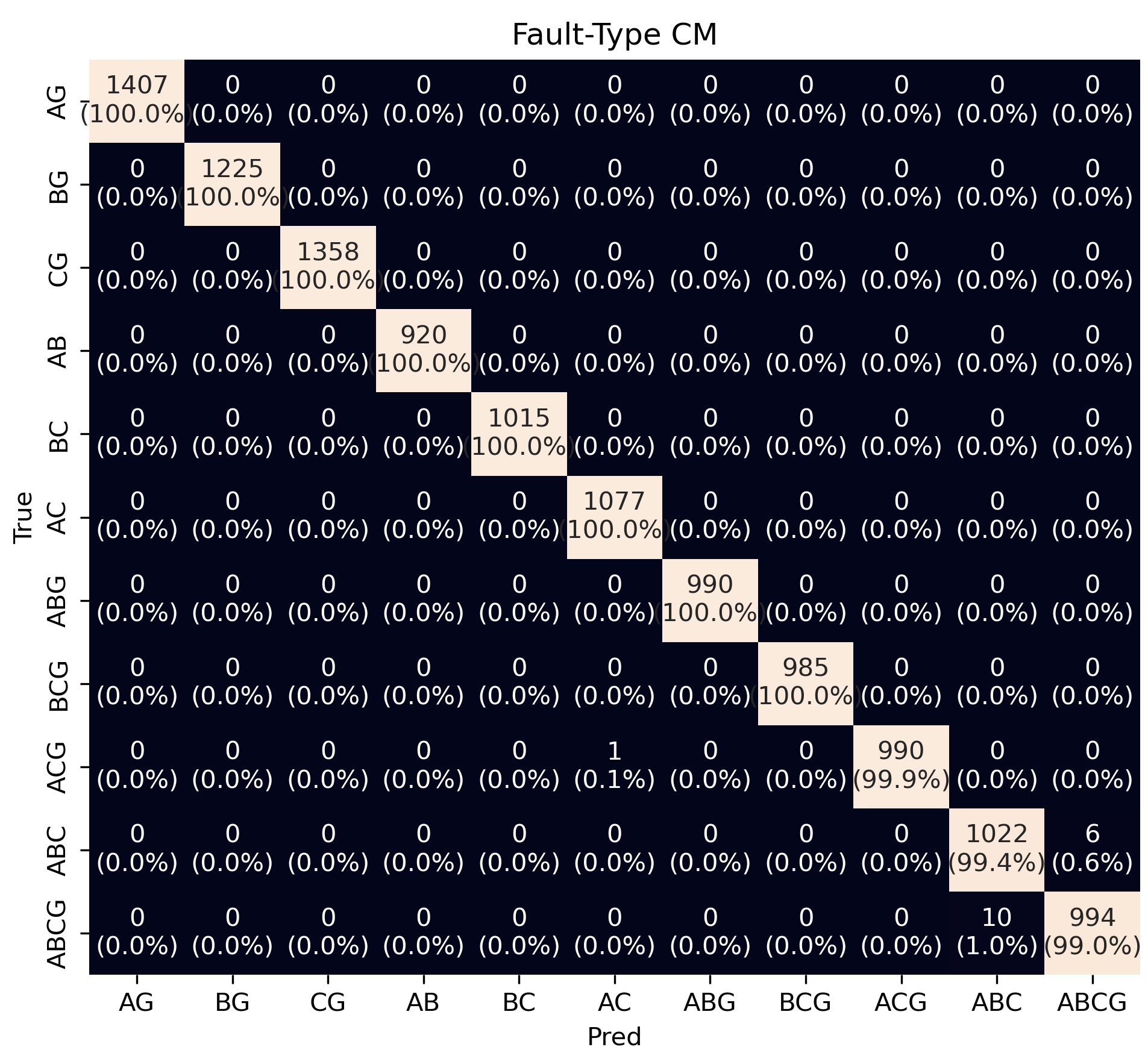}
  \caption{Confusion matrix of fault classification for IEEE-123 Bus System with PMUs selected using FSNR-SVM.}
  \label{fig:IEEE-123-confusion-mat}
\end{figure}

For the IEEE-123 bus system (Table \ref{tab:perf}), the best results for faulted line and fault type identification are obtained with PMUs placed using the FSNR-SVM algorithm. FSNR with Random Forest yields nearly identical results, as both select buses in the same neighborhood. Most line misclassifications occur between single-phase lines sharing a common bus. The confusion matrix in Fig. \ref{fig:IEEE-123-confusion-mat} shows that the misclassifications in identifying the fault types are only between ABC and ABCG three-phase faults.    

\begin{table*}[!htb]
\centering
\footnotesize
\caption{Classification performance for fault location and fault type}
\label{tab:perf}
\begin{tabular}{@{}l l  cccccc  ccccc@{}}
\toprule
\textbf{\multirow{2}{*}{Test System}} 
& \textbf{\multirow{2}{*}{\shortstack{PMU Placement \\ Method}}}
  & \multicolumn{5}{c}{\textbf{Fault Location}}
  & \multicolumn{5}{c}{\textbf{Fault Type}} \\
\cmidrule(lr){3-7} \cmidrule(lr){8-12}
 & 
  & Acc. & Prec. & Rec. & F1 & Spec.
  & Acc. & Prec. & Rec. & F1 & Spec. \\
\midrule
\multirow{4}{*}{IEEE-34}
 & FSNR–SVM       &\blue{0.96411} &\blue{0.96463} &\blue{0.96438} &\blue{0.9645}	&\blue{0.9989}	&0.99389 &0.99364 &0.9938 &0.99372 &0.99939
 \\
& FS-SVM &0.96122 &0.96223 &0.96298 &0.96261 &0.99882 &0.99267 &0.99246 &0.99243&0.99244&0.99927 
\\
 
  & FSNR–RF         &0.93133	&0.9275	&0.9254	&0.92645	&0.99791	&0.99144	&0.99127	&0.99111	&0.99119	&0.99915

                   \\
  & Corr.\ Distance &0.94711	&0.94422	&0.9473	&0.94576	&0.99839	&0.99356	&{0.99342}	&{0.9933}	&0.99336	&{0.99936}
 
                   \\
  & Y-Mat Based     &0.95433	&0.95402	&0.95608	&0.95505	&0.99861	&\blue{0.99489}	&\blue{0.99466}	&\blue{0.99485}	&\blue{0.99476}	&\blue{0.99949}

 \\
\midrule
\multirow{4}{*}{IEEE-123}
  & FSNR–SVM  & \blue{0.94075}	&0.83615	&0.85429	&0.84512	&\blue{0.99949}	&\blue{0.99858}	&\blue{0.99849}	&\blue{0.99847}	&\blue{0.99848}	&\blue{0.99986}
\\
  & FSNR–RF         &0.93442	&\blue{0.83794}	&\blue{0.85755}	&\blue{0.84763}	&0.99944	&0.99825	&0.99812	&0.99812	&0.99812	&0.99983

  \\
  & Corr.\ Distance & 0.83967 & 0.75599	 & 0.77495 & 0.76535 & 0.99862	 
                   & 0.99825	 & 0.99813 & 0.99811 & 0.99812 & 0.99983 \\
  & Y-Mat Based     &0.88983	&0.78883	&0.7931	&0.79096	&0.99906	&0.99817	&0.99803	&0.99803	&0.99803	&0.99982 \\

\bottomrule
\end{tabular}
\end{table*}

%---------------------------------
\vspace{1 em}
\section{Conclusion}
In this paper, a data driven Forward Search algorithm with Neighborhood Refinement is proposed to help in selecting the best locations for placing PMUs to record pre-fault and during fault voltage and current values that can be used with deep neural network architectures. We use a cross-validated, SVM based classifier that uses difference of pre-fault and during fault measurements to select the most informative PMU locations within a PDN. We include voltage and current measurements from the substation and only voltage measurements from other possible PMU locations. The substation data is always included in the search algorithm. With neighborhood refinement, the impact of globally suboptimal selection can be mitigated. SVM based classifiers have been shown to be effective when data is limited and, compared to other machine learning techniques, are computationally less expensive. 

Using a 1D CNN model for evaluation, we compare our proposed FSNR-SVM algorithm with other techniques. The results demonstrate that our method effectively identifies the number and placement of PMUs that consistently deliver the best overall performance.

\FloatBarrier
\bibliographystyle{IEEEtran}  
\bibliography{bibliography}

\end{document}